\documentclass[twocolumn,preprintnumbers,amsmath,amssymb]{revtex4}

\usepackage[utf8]{inputenc}
\usepackage{hyperref}
\usepackage{graphicx}
\usepackage{amsmath}
\usepackage{bm}
\usepackage{amssymb}

\usepackage{color}

\newcommand{\tr}[1]{{\textrm{#1}}}

\begin{document}

\title{Giant Stark splitting of an exciton in bilayer MoS$_\mathbf{2}$}

\author{Nadine Leisgang$^{1*}$}
\author{Shivangi Shree$^{2*}$}
\author{Ioannis Paradisanos$^{2*}$}
\author{Lukas Sponfeldner$^{1*}$} 
\author{Cedric Robert$^2$}
\author{Delphine Lagarde$^2$}
\author{Andrea Balocchi$^2$}
\author{Kenji Watanabe$^3$}
\author{Takashi Taniguchi$^4$}
\author{Xavier Marie$^2$}
\author{Richard J. Warburton$^1$}
\author{Iann C. Gerber$^2$}
\author{Bernhard Urbaszek$^2$}

\affiliation{\small$^1$Department of Physics, University of Basel, Basel, Switzerland}
\affiliation{\small$^2$Universit\'e de Toulouse, INSA-CNRS-UPS, LPCNO, 135 Avenue Rangueil, 31077 Toulouse, France}
\affiliation{\small$^3$Research Center for Functional Materials, 
National Institute for Materials Science, 1-1 Namiki Tsukuba, Ibaraki
305-0044, Japan}
\affiliation{\small$^4$International Center for Materials Anorthite, 
National Institute for Materials Science, 1-1 Namiki Tsukuba Ibaraki
305-0044, Japan}

\maketitle

\textbf{
Transition metal dichalcogenides (TMDs) constitute a versatile platform for atomically thin optoelectronics devices and spin-valley memory applications. In monolayer TMDs optical absorption is strong, but the transition energy can not be tuned as the neutral exciton has essentially no out-of-plane static electric dipole \cite{roch2018quantum,verzhbitskiy2019suppressed}. In contrast, interlayer exciton transitions in heterobilayers are widely tunable in applied electric fields, however their coupling to light is significantly reduced. In this work, we show tuning over 120~meV of interlayer excitons with high oscillator strength in bilayer MoS$_{2}$ due to the quantum confined Stark effect \cite{miller1984band}. We optically probe the interaction between intra- and interlayer excitons as they are energetically tuned into resonance. Interlayer excitons interact strongly with intralayer B-excitons as demonstrated by a clear avoided crossing, whereas the interaction with intralayer A-excitons is significantly weaker. Our observations are supported by density functional theory (DFT) calculations including excitonic effects. In MoS$_2$ trilayers, our experiments uncover two types of interlayer excitons with and without in-built electric dipoles, respectively. Highly tunable excitonic transitions with large in-built dipoles and oscillator strengths will result in strong exciton-exciton interactions and therefore hold great promise for non-linear optics with polaritons.}

\begin{figure*}[t]
\centering
\includegraphics[width=153mm]{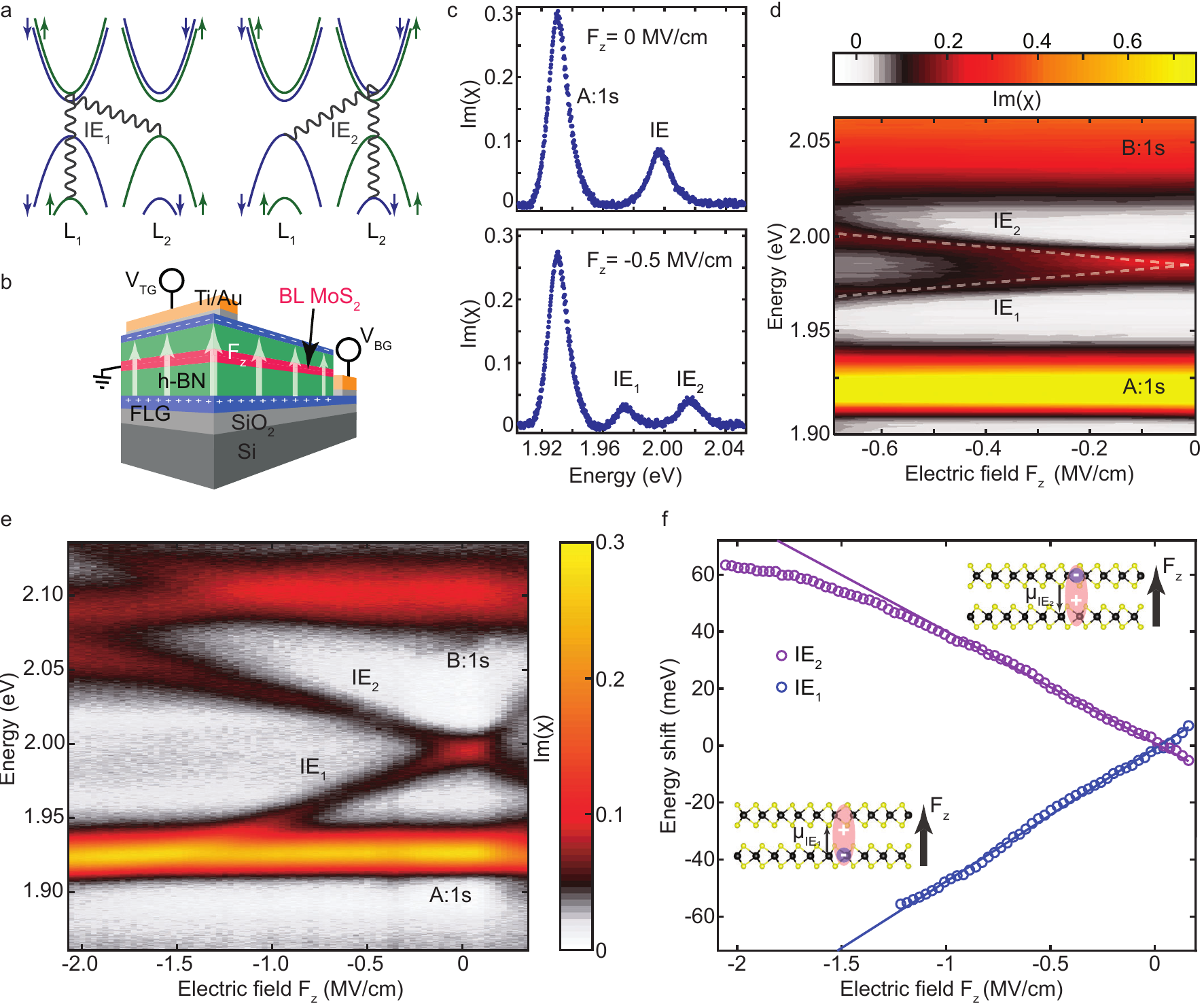} 
\caption{\textbf{MoS$_\mathbf{2}$ bilayer van der Waals heterostructure (vdWH) in an applied electric field at $\bm{T=4}$~K}. \textbf{a, }Schematic of interlayer exciton configurations IE$_1$ and IE$_2$ with a strong B-intralayer exciton component at $F_z=0$~MV/cm according to analysis in \cite{gerber2019interlayer}. \textbf{b,} Three-dimensional schematic of device 1, consisting of a MoS$_{2}$ homobilayer encapsulated between two hexagonal boron nitride (hBN) flakes. Few-layer graphene (FLG) serves as top and bottom gate, while a direct Ti/Au contact to the MoS$_{2}$ is used to ground the bilayer. Voltages to the top and bottom gates ($V_\tr{TG}$ and $V_\tr{BG}$) are applied to create a uniform electric field $F_z$ across the device. \textbf{c,} Typical absorption spectra recorded without ($F_z=0$~MV/cm) and with ($F_z=-0.5$~MV/cm) an applied electric field, extracted from \textbf{e}. \textbf{d,} Colourmap of the absorption spectra of bilayer MoS$_2$ vdWH (device 2). Stark shift of the interlayer excitons at small electric fields. The intralayer A- and B-excitons (A:1s and B:1s) and the two branches of the interlayer A-resonances (IE$_1$ and IE$_2$) are labelled. \textbf{e,} Colourmap of the absorption spectra of device 1 as a function of the electric field $F_z$ applied perpendicular to the vdWH. \textbf{f,} Stark shift of the interlayer A-excitons as a function of $F_z$, extracted from the spectra in panel \textbf{e}. The solid blue and purple lines are linear fits to the experimental data points at small/moderate electric fields ($F_z=0.1$~MV/cm to $F_z=-1$~MV/cm). Insets show schematics of interlayer excitons in homobilayer MoS$_2$. An electron localised in one layer interacts with a hybridised hole state to form an interlayer exciton. The direction of the dipole moment depends on the location of the electron, either in the bottom ($\mu_{\tr{IE}_1}$) or top ($\mu_{\tr{IE}_2}$) layer.}
\label{fig1}
\end{figure*}

The optical properties of TMDs, such as MoS$_{2}$ and WSe$_{2}$, are governed by excitons, Coulomb bound electron-hole pairs \cite{Chernikov:2014a,Wang:2018a}. High quality van der Waals heterostructures show close-to-unity, gate-tunable reflectivity of a single MoSe$_2$ layer \cite{Scuri:2018a,PhysRevLett.120.037401}, variation of the transition energies of interlayer excitons over a broad wavelength range in heterobilayers \cite{rivera2015observation,joe2019electrically} and valley polarised exciton currents \cite{unuchek2019valley}.\\
\indent In heterobilayers, the interlayer exciton is formed with the electron either in the top or in the bottom layer depending on the initial stacking \cite{rivera2015observation}. Reports on interlayer excitons in heterobilayers rely mostly on photoluminescence emission \cite{Nagler2017,Forg2019cavity-control,rivera2015observation,unuchek2019valley,joe2019electrically} as interlayer absorption is very weak.
In 2H stacked MoS$_2$ homobilayers, the situation is different: First, a strong feature in absorption up to room temperature has been observed in earlier studies on MoS$_2$ bilayers \cite{slobodeniuk2019fine,calman2018indirect,gerber2019interlayer,niehues2019interlayer,carrascoso2019biaxial} and interpreted as an interlayer exciton, as theoretically predicted by Deilmann et al.\ \cite{deilmann2018interlayer}. It was proposed that the strong oscillator strength of the interlayer exciton originates from a strong admixture with the B-intralayer transition (see scheme in Fig.~\ref{fig1}a and Supplementary Fig.~1, and discussion in previous calculations \cite{deilmann2018interlayer,gerber2019interlayer}). Second, in principle, two energetically degenerate interlayer excitons can form with the electron residing in either the top or the bottom layer \cite{PhysRevLett.123.117702}, whereas the hole is delocalised \cite{gong2013magnetoelectric}. \\
\begin{figure*}[t]
\centering
\includegraphics[width=153mm]{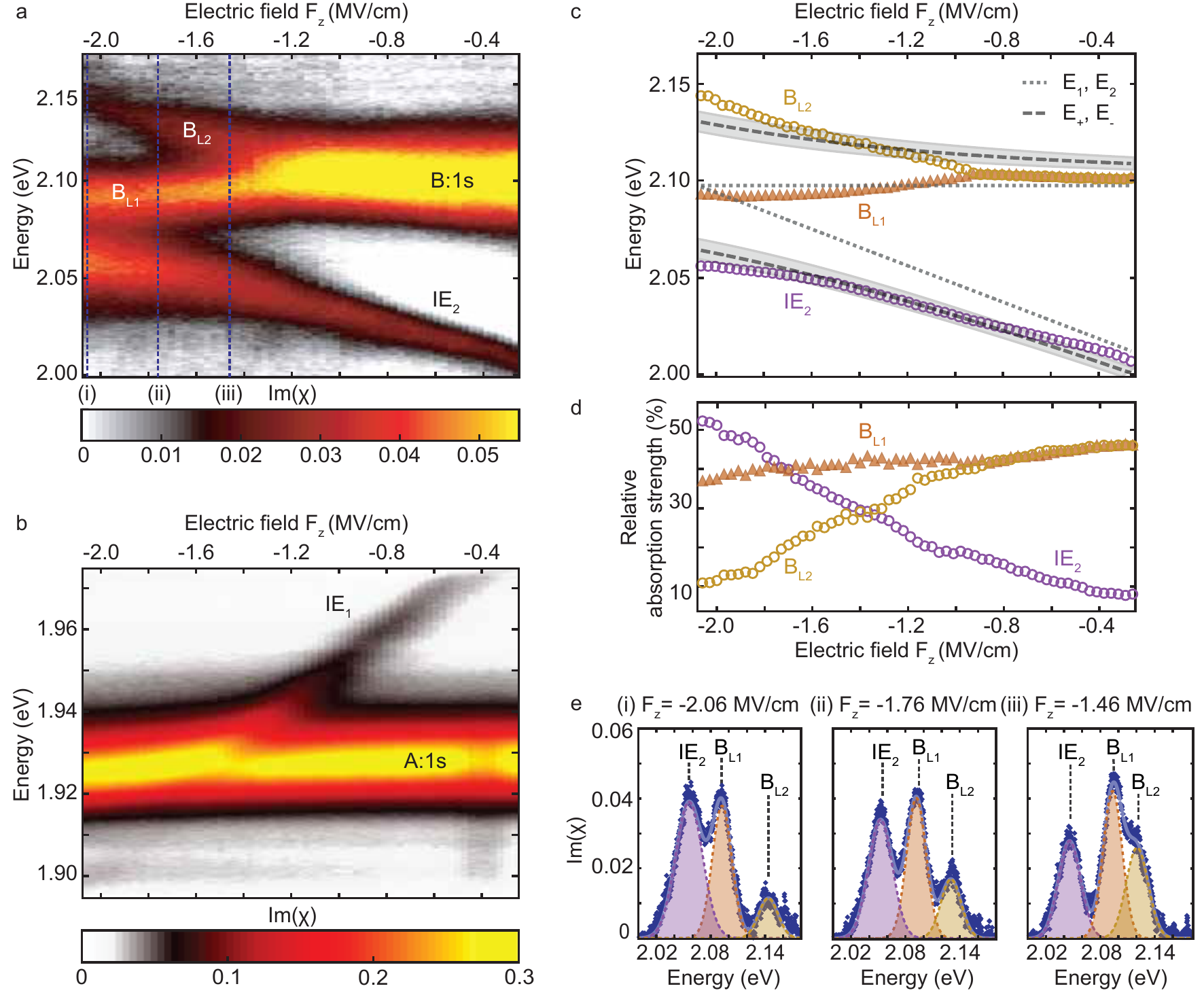} 
\caption{\textbf{Electric field dependence in a magnetic field ($\bm{B_z=+9}$~T).} \textbf{a,} Colourmap of the absorption spectra ($\sigma^+$-polarisation) centred around the intralayer B-exciton.  B$_{\tr{L1}}$ and B$_{\tr{L2}}$ are the intralayer excitons residing in layer 1 (L$_1$) and 2  (L$_2$), respectively. Dashed vertical lines show spectra reproduced in panel \textbf{e}. Experiments were performed at $\bm{B_z=+9}$~T for greater mechanical stability of the set-up. \textbf{b,} Colourmap of the absorption spectra ($\sigma^+$-polarisation) centred around the intralayer A-exciton, showing a very small avoided crossing with IE$_1$. \textbf{c,} Peak positions versus applied electric field extracted from spectra shown in panel \textbf{a}, the transitions IE$_2$ and B$_{\tr{L2}}$ show a clear avoided crossing. The dotted lines show the peak energy evolution expected without coupling. The dashed lines are a fit (see Supplementary Section~V) using a coupling energy $W$ of $33\pm 5~$meV; the shaded area corresponds to the uncertainty in $W$. \textbf{d,} Evolution of the integrated absorption strength of IE$_2$, B$_{\tr{L2}}$ and B$_{\tr{L1}}$ confirming the mixing of IE$_2$ and B$_{\tr{L2}}$; here 100\% corresponds to the sum of these three transitions. \textbf{e,} Spectra from data of panel \textbf{a,} with the three-peak fit that  determines the transition energies and the relative absorption strengths (plotted in panels \textbf{c,d})}
\label{fig2}
\end{figure*}
\indent In this Letter, we investigate excitons in bilayer MoS$_2$ with both strong light-matter interaction and high tunability in external electric fields.
Our experiments focus on momentum-direct intralayer and interlayer excitons originating from valence and conduction bands around the $K$-point. 
We integrate MoS$_{2}$ bilayers in devices (1 and 2) with top and bottom gates for applying a static out-of-plane electric field $F_z$ (see Method Section and Supplementary Section~II,~IV). Fig.~\ref{fig1}e shows typical absorption spectra as a function of the electric field $F_z$, recorded on device 1 (see Fig.~\ref{fig1}b for a three-dimensional schematic of the device). Three prominent transitions can be clearly identified at zero electric field ($F_z=0$~MV/cm): the intralayer A- and B-excitons (A:1s and B:1s) near 1.93~eV and 2.10~eV, respectively, and the interlayer A-exciton (IE) at 2.00~eV \cite{gerber2019interlayer}. On applying an external electric field $F_z$, the IE splits into two well-separated branches, as seen in Fig.~\ref{fig1}c for $F_z=-0.5$~MV/cm. Using the area under the peaks in Fig.~\ref{fig1}c as a rough measure of the relative absorption strength (we take A:1s as 100\%), we see that at $F_z=0$~MV/cm the IE is about 30\% compared to the intralayer A:1s. Remarkably, at finite electric fields $F_z=-0.5$~MV/cm, the absorption strength does not vanish but remains rather strong, with a combined 24\% from IE$_1$+IE$_2$ relative to the A:1s. Our experiments clearly show that the absorption peak IE, initially at 2.00~eV, corresponds to interlayer exciton resonances with out-of-plane oriented electric dipoles: The carriers clearly do not reside within the same layer. For the intralayer excitons on the other hand, the energy shift with applied electric field is negligible, as in the case for excitons in monolayers \cite{roch2018quantum,verzhbitskiy2019suppressed}. As $F_z$ is increased, the energy difference between the IE$_1$ and IE$_2$ states reaches a value of $\sim 120$~meV, covering a wide spectral range, spanning the energy range between the intralayer A- and B-excitons. \\
\indent For small to moderate electric fields, before significant interaction between the interlayer states and the A- and B-excitons, we observe a linear energy shift with $F_z$ for both peaks, IE$_1$ and IE$_2$, suggesting a first-order Stark shift caused by the static electric dipole moments across the MoS$_2$ bilayer (Fig.~\ref{fig1}d,e). In Fig.~\ref{fig1}f we plot the transition energies extracted from Fig.~\ref{fig1}e as a function of the applied field $F_z$ and perform a linear fit. We extract large dipole moments of $\mu_{\tr{IE}_1}=(0.47~\pm~0.01)$~$e\cdot$nm and $\mu_{\tr{IE}_2}=(-0.39~\pm~0.01)$~$e\cdot$nm with $e$ being the electron charge. Applying higher electric fields in our experiment, we discover that the shifts deviate from a simple linear Stark shift, reflecting, as discussed below, very different interactions of the interlayer excitons with the A- and B-intralayer excitons. For the analysis of device 2, shown in Fig.~\ref{fig1}d, we extract a dipole value with a lower bound of about $0.3$~$e\cdot$nm. We confirm the interlayer character of the IE$_{1,2}$ excitons also in magneto-optics: Interlayer excitons show a larger $g$-factor with opposite sign compared to intralayer excitons \cite{arora2017interlayer} (see Supplementary Section~VI for our measurements)
\begin{figure*}[t]
\centering
\includegraphics[width=153mm]{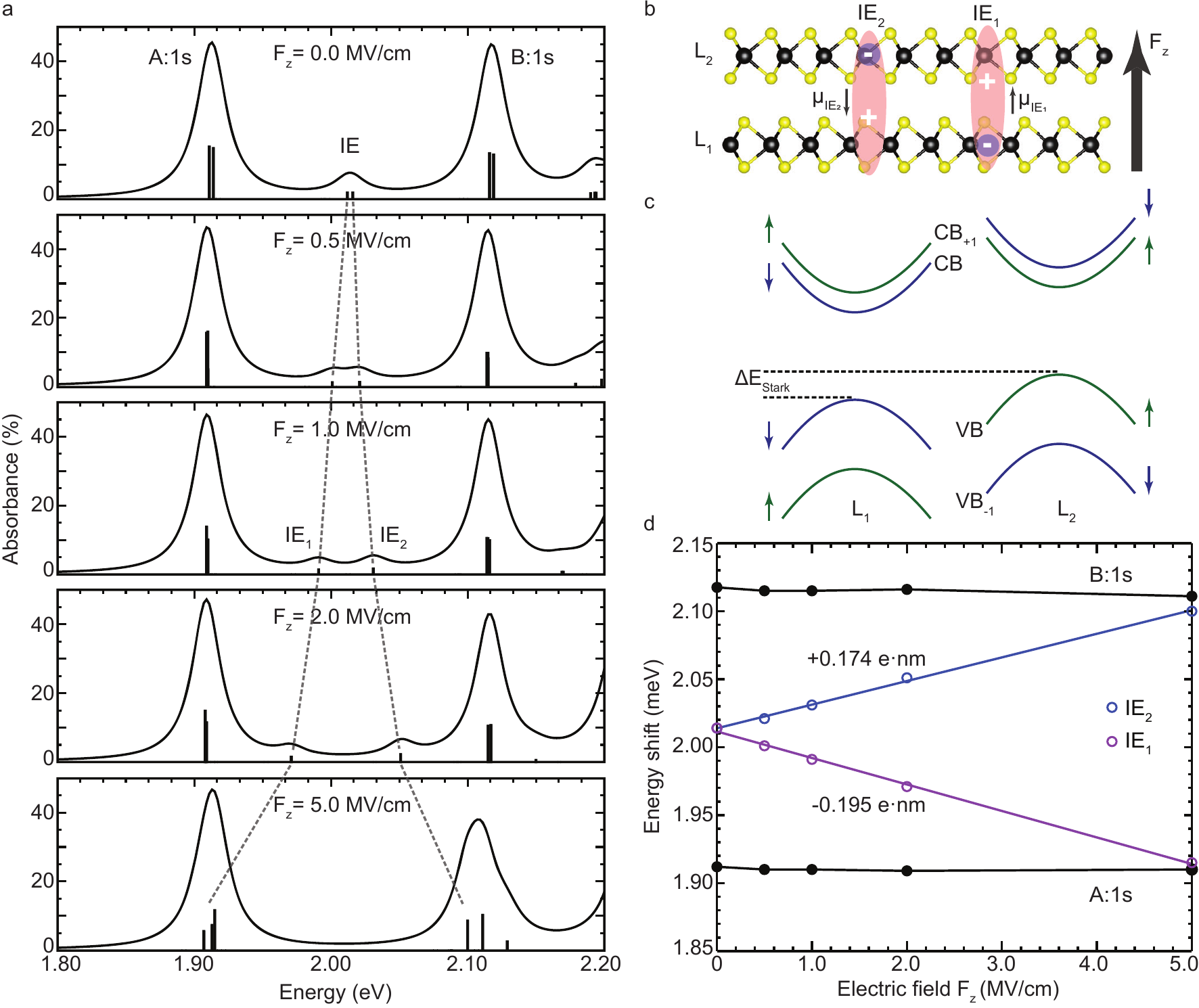}
\caption{\textbf{Beyond-DFT calculations of the electric field effects on the band structure and excitonic properties of 2H MoS$\mathbf{2}$ bilayer}. \textbf{a,} Absorbtion spectra as a function of the applied electric field $F_z$. Our DFT-$GW$-BSE calculations determine both energy position and absorption strength corresponding to exciton transitions as shown with vertical black lines. We estimate the numerical precision of our calculations to be of the order of $\pm5~$meV (see computational settings in Supplementary Section~VII) such that any splitting below this value does not have any physical significance. For comparison with the experiment, we introduce an artificial broadening of the order of 10 meV for each transition. This results in the spectra shown with solid black lines; the three main transitions A:1s (intralayer A), IE (interlayer) and B:1s (intralayer B) are labelled. \textbf{b,} Structural model of the 2H bilayer stacking and electric field direction. \textbf{c,} Schematic of the band structure modification due to the application of $F_z$ showing a global shift, denoted $\Delta E_{\tr{Stark}}$, between the distinct band structures of the two layers (see Supplementary Section~VII for detailed results on $GW$ band structure calculations). \textbf{d,} Evolution of the excitonic peak positions with respect to the electric field, and the corresponding dipole moment estimates for IE$_1$ and IE$_2$ excitons.\label{fig3}}
\end{figure*}
In terms of the magnitude, the large extracted dipole moments are similar to results on interlayer excitons in MoSe$_2$/WSe$_2$ \textit{heterobilayers} as determined by photoluminescence \cite{unuchek2019valley,doi:10.1002/pssb.201900308,rivera2015observation,joe2019electrically}. 
For comparison with other \textit{homobilayer} systems, interlayer excitons have very different characteristics depending on the TMD material \cite{wu2013electrical,Jones:2014a}. For WSe$_2$ bilayers, reports on interlayer excitons focus on transitions indirect in momentum space linked to the indirect band gap energetically below the direct $K$-$K$ transitions \cite{Jessi2018,wang2017electrical}. In the case of MoSe$_2$, $K$-$K$ interlayer excitons are observed with similarities to the case of MoS$_2$, but with significantly lower oscillator strength \cite{PhysRevB.97.241404,shimazaki2019moir,sung2020broken}. This is mainly due to the larger valence band spin-orbit splitting of MoSe$_2$ compared to MoS$_2$, which makes hole delocalisation over both layers and hence the formation of interlayer excitons less favourable \cite{gong2013magnetoelectric}. Signatures of interlayer excitons in bulk 2H MoSe$_2$ have been reported, but the same study did not find $K$-$K$ interlayer exciton signatures for bulk 2H WSe$_2$ \cite{arora2018valley}. \\
\indent Due to the very large tunability shown in Figs.~\ref{fig1}d,e for the interlayer transition energy, we are able to tune the interlayer excitons into resonance with the A- and B-intralayer excitons. The interactions are very different. This can be directly seen by comparing Fig.~\ref{fig2}a (IE$_2$ $\leftrightarrow$ B) with \ref{fig2}b (IE$_1$ $\leftrightarrow$ A). In Fig.~\ref{fig2}b, the IE$_1$ transition merges with the A-exciton line at around -1.3~MV/cm; at the highest electric fields in these experiments, the absorption contrast of IE$_1$ decreases (a consequence of the large change to the dielectric constant induced by the strong A-exciton), falling below the noise. This points to a small avoided crossing of the weak IE$_1$ exciton with the strong A-exciton. On the one hand, the IE-A coupling is clearly less than the linewidth of the A-exciton – if this were not the case then a clear avoided crossing would be visible. On the other hand, the coupling is not zero - there is clearly a change in the dispersion of the A-exciton at the electric fields where the IE and A-excitons are close in energy. This point can best be probed on samples with a smaller linewidth. Our estimate for the coupling of the IE-A excitons on the present sample is $5 \pm 3$~meV.\\
\indent In contrast, tuning the upper interlayer branch IE$_2$ energetically close to resonance with the B-exciton leads to a clear avoided crossing, see data in Fig.~\ref{fig2}a and extracted transition energies in Fig.~\ref{fig2}c. For the B-exciton, we can distinguish two resonances associated to intralayer excitons in the two different layers L$_1$ and L$_2$, labelled as B$_{\tr{L1}}$ and B$_{\tr{L2}}$. The transition B$_{\tr{L1}}$ does not share any state with IE$_2$ as evidenced by a nearly constant transition energy and integrated absorption strength in this electric field range, see  Fig.~\ref{fig2}d,e. 
Importantly, as IE$_2$ has a strong B$_{\tr{L2}}$-intralayer component, see scheme in Fig.~\ref{fig1}a, these states interact strongly. 
There are two experimental signatures for strong coupling between B$_{\tr{L2}}$  and IE$_2$: First, we observe a clear avoided crossing in Fig.~\ref{fig2}c on plotting the transition energies as a function of the applied field. Second, B$_{\tr{L2}}$ and IE$_2$ exchange strength: B$_{\tr{L2}}$ is initially strong but then weakens; IE$_2$ is initially weak but then strengthens, becoming the dominant exciton at large electric fields. This can be clearly seen on plotting the evolution of the absorption strength in Fig.~\ref{fig2}d. By using a simple two-level model for the two transitions B$_{\tr{L2}}$  and IE$_2$ coupled by an interaction $W$, we can fit the data (dashed lines in Fig.~\ref{fig2}c) and extract a coupling energy $W=33~$meV$\pm5$~meV (see Supplementary Section~V for details).\\
\indent The main observations of the experiments on devices 1 and 2 are (i) the splitting and very large energy shifts of the IE transitions in an applied electric field, (ii) anticrossing of the IE$_2$ with the B-intralayer exciton, and (iii) small avoided crossing of the IE$_1$ with the A-intralayer exciton. The comparatively large Stark shift and the strong enough oscillator strength for absorption experiments have been initially discussed by Deilmann et al.\ \cite{deilmann2018interlayer} using $GW$+BSE calculations. Our target here, also using $GW$+BSE calculations, is to develop a semi-quantitative understanding for the observed effects (i) and (ii) by calculating the mixing of intra- and interlayer exciton components as a function of $F_z$.
Our approach is based on the inclusion of the applied electric field as a perturbation in the band structure calculations, similar to the approach in  \cite{deilmann2018interlayer,fan2016valence,liu2012tuning} (see Supplementary Section~VII for computational details). Please note that we have only considered freestanding 2H MoS$_2$ bilayers in our calculations, i.e.\ placed in vacuum, as we aim to qualitatively reproduce the main trends. When applying an electric field, we observe a global shift for L$_1$ of the relevant conduction and valence bands at the $K$-point down in energy with respect to L$_2$, marked as $\Delta E_{\text{Stark}}$ (see Fig.~\ref{fig3}b for the electric field configuration and Fig.~\ref{fig3}c for a band structure schematic). Please also see Supplementary Section~VII for a more realistic $GW$ picture of this effect on the band structure. 
From the sketch in Fig.~\ref{fig3}c, it is thus clear that the IE transitions will split in energy: Transitions involving the L$_2$ valence bands and the conduction bands in L$_1$ will lower in energy, whereas transitions involving the L$_1$ valence bands and L$_2$ conduction bands will increase in energy. \\
\indent After including excitonic effects, we are able to calculate the absorption (in Fig.~\ref{fig3}a) which looks very similar to our measurements at small electric fields (see Fig.~\ref{fig1}c). Fig.~\ref{fig3}d demonstrates how the IE transition energy changes, roughly linearly, with the applied electric field. At large electric field values of 5~MV/cm, we see two main groups of transitions: At the low energy side, the IE$_1$ close to the A-exciton energy and, at the high energy side, the IE$_2$ close to the B-exciton as in \cite{deilmann2018interlayer}. Our calculations show relative IE oscillator strengths versus A:1s ones relatively close to experiments, despite neglecting the environmental dielectric constant variations. There are essentially no transitions in the energy range in-between. We refrain in our calculations to comment on any higher electric field values as our initial assumption to treat the electric field as a perturbation, in comparison with other energy splittings in the band structure, reaches its limits.\\
\indent The exact nature and mixing of the absorption peaks, calculated in Fig.~\ref{fig3}a, contain information on the evolution of the exciton states with electric field (see Supplementary Section~VII for summary tables). We can use this knowledge to analyse the results in Fig.~\ref{fig2}a and b in particular: For finite electric fields, the intralayer A-exciton contains a very small IE component that remains small as the field increases. Our theory indicates very little mixing between the A-intralayer exciton and the IE, consistent with the experiments in Fig.~\ref{fig2}b. Concerning interactions with the intralayer B-exciton, the IE exciton, with a delocalised hole, is mixed with the intralayer B-exciton as they share the same valence states. This was suggested in \cite{gerber2019interlayer} to explain the surprisingly high oscillator strength of the IE transitions. In our calculations, we see that the mixing between the higher energy IE branch and the B-exciton becomes stronger as the electric field amplitude is increased (see Supplementary Section~VII for a detailed description of each component). We suggest that this clear admixture of the IE-exciton with the B-exciton in our calculations is the origin of the observed avoided crossing between IE$_2$ and B$_{\tr{L2}}$ shown in Fig.~\ref{fig2}a and c, as well as the enhancement of the absorption strength of IE$_2$, shown in Fig.~\ref{fig2}d. Our $GW$+BSE calculations therefore capture the main experimental findings, and aid our understanding of the interactions of the IE with the intralayer A- and B-exciton resonances, and the observed IE Stark shift, predicted by \cite{deilmann2018interlayer}. \\
\begin{figure*}[t]
\centering
\includegraphics[width=153mm]{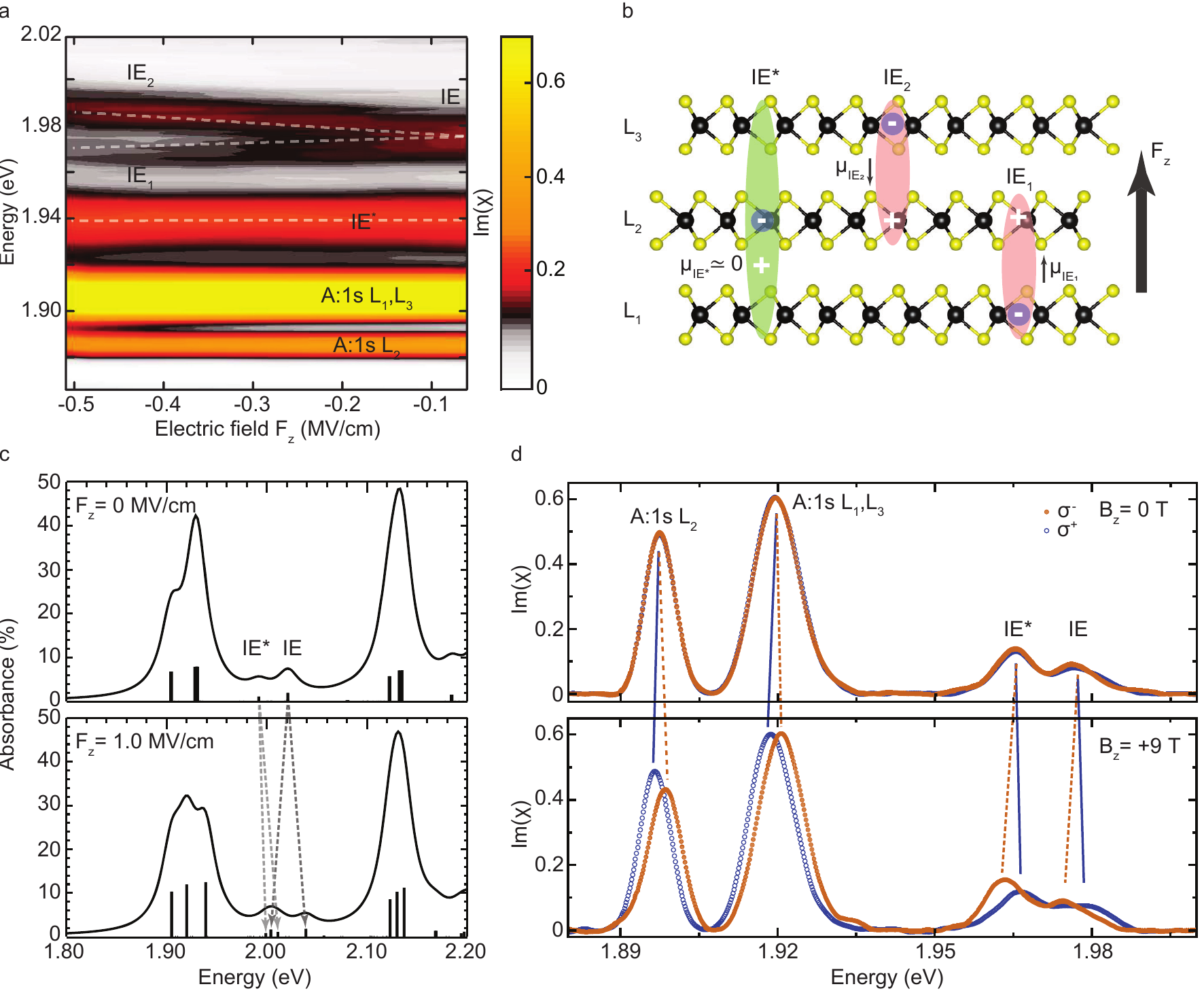} 
\caption{\textbf{MoS$_\mathbf{2}$ trilayer in an applied electric field.} \textbf{a,} Colourmap of the absorption spectra in device 2. \textbf{b,} Schematic of 2H stacked trilayer studied with scheme representing the microscopic origin of IE and IE$^*$. \textbf{c,} Theoretical absorption spectra of 2H MoS$_2$ freestanding trilayer without (top) and with (bottom) electric field. See Supplementary Section~VII for a detailed discussion of the trilayer band structure and the effect of the electric field on it. The Stark shifts of interlayer excitons IE and IE$^*$ are indicated by arrows. \textbf{d,} Magneto-optics on device 3 with small Zeeman splitting for intralayer excitons (A:1s) and large Zeeman splitting for interlayer excitons IE and IE$^*$ in magnetic fields of $B_z=$+9~T perpendicular to the monolayer plane.}
\label{fig4}
\end{figure*}
\indent We apply our combined approach using experiments and theory to uncover and manipulate novel exciton species in a more complex system, namely MoS$_2$ homotrilayers. In contrast to inversion symmetric MoS$_2$ bilayers, trilayers could be advantageous in non-linear optics due to the broken inversion symmetry that gives rise to a non-vanishing second-order non-linear susceptibility. In the trilayer system, we uncover additional types of interlayer excitons as compared to the bilayer in our experiment:
At zero electric field, two different interlayer transitions appear in absorption, labelled IE$^*$ and IE in Fig.~\ref{fig4}a. We show absorption measurements as a function of the electric field and make a surprising observation: Whereas IE splits into two branches (IE$_1$ and IE$_2$) as for the bilayer studies, IE$^*$ does not show any measurable splitting and hence, a negligible or, at least, a very small in-built electric dipole moment. Interestingly, we make the same observation in our $GW$+BSE calculations, shown in Fig.~\ref{fig4}c. We can explain the different behaviour in applied electric fields by analysing the microscopic origin of IE and IE$^*$ in our calculations (see Supplementary Section~VII). The schematic in Fig.~\ref{fig4}b shows these excitons to have very distinct characteristics: For the IE, two degenerate states form at zero electric field, i.e.\ IE$_1$ and IE$_2$. For IE$_1$, the electron is localised in the bottom layer and the hole is delocalised between the middle and bottom layers. For IE$_2$, the electron is localised in the top layer and the hole is delocalised between the middle and top layers. As such, IE$_1$ and IE$_2$ have a finite dipole moment, similar to the interlayer excitons in bilayers. 
We extract a dipole moment of $|\mu| \approx 0.15$~$e\cdot$nm. This smaller value as compared to the bilayer can have its origin in the different dielectric environment and a more localised hole wavefunction. Admixture of intralayer with interlayer excitons does not change significantly over the investigated electric field range (see Supplementary Table~II).
For the IE$^*$ at slightly lower energy, the situation is different compared to IE: The electron is localised in the middle layer and the hole is delocalised over all three layers. This results in a negligible in-built electric dipole moment of IE$^*$ which translates into a non-resolvable Stark splitting of the transition in our measurements.
The absence of a clear Stark shift might put the interlayer nature of IE$^*$ into question. To answer this, we have performed magneto-optics, shown in Fig.~\ref{fig4}d. We observe for the intralayer excitons a Zeeman splitting of the order of 2~meV for $B_z=+9$~T and for IE and IE$^*$, a Zeeman splitting of the order of 4~meV, with opposite sign compared to the intralayer transitions. Therefore, we confirm the interlayer character of these peaks. These results lead to an unusual situation for interlayer excitons: We observe a strong splitting in magnetic fields for IE$^*$, but a very small Stark shift in electric fields. \\
\indent The transition energy of interlayer excitons in bilayer MoS$_2$ can be tuned over 120~meV, more than 10 times their linewidth.
The interaction of the blue-shifted interlayer exciton with the intralayer B-exciton shows clear signatures of strong coupling. In contrast, the interaction of the red-shifted interlayer exciton with the intralayer A-exciton shows only a weak coupling. This allows us to conclude that the interlayer exciton has a strong B-exciton component in its wave function but a much smaller A-exciton component. We expect that at even higher fields than those in these experiments, the interlayer exciton will lie lower in energy than the A-exciton and will recover its absorption strength once it is sufficiently red-detuned from the A-exciton. This will represent an advantageous scenario for applications: The ground-state exciton is long-lived and possesses both a large in-built electric dipole moment and a strong absorption.\\
\indent For optoelectronics, the highly tunable excitons with an in-built dipole found here in MoS$_2$ bilayers are promising for exploring coupling to optical cavity modes, where excitons and photons can couple to form polaritons \cite{Cristofolini704,Schneider2018a}. In III-V semiconductors very recently optical non-linearities at a single polariton level have been detected \cite{munoz2019emergence,delteil2019towards}. Building on these promising proof-of-principle experiments further work on excitonic systems with stronger exciton-exciton interactions is desirable. This is the case for interlayer excitons in general in TMDs with in-built static dipole. The interlayer excitons in the homobilayers investigated here have the crucial advantage of a high oscillator strength visible in absorption. \\

\noindent \textbf{Methods} \\
We performed experiments as a function of the out-of-plane electric field in 2H stacked bilayer MoS$_2$ independently in two different research laboratories using devices 1 and 2, with a geometry as depicted in Fig.~\ref{fig1}b: Two hexagonal boron nitride (hBN) flakes are used as dielectric spacers, and top and bottom few-layer graphene (FLG) act as transparent electrodes (see Supplementary Section~II for a description of the fabrication process and optical images of the devices). In device 1, a direct Ti/Au contact is added to the MoS$_2$ to operate the system in a dual-gate device scheme by grounding the bilayer, allowing independent control of the applied electric field and the carrier concentration. Applying a DC voltage to the top and bottom gate ($V_\tr{TG}$ and $V_\tr{BG}$) creates a uniform electric field in the MoS$_2$, oriented perpendicular to the bilayer (see Supplementary Section~IV). The optical reflectivity was measured at low temperature ($T=4$~K) using a home-built confocal microscope and a weak, incoherent light source (see Supplementary Section~III). The imaginary part of the optical susceptibility Im($\chi$), a measure of the absorption, was deduced from the differential reflectivity signal $\Delta R/R_0$, $\Delta R=R-R_0$, using the Kramers-Kronig relation, where $R$ is the reflectivity spectrum obtained on the MoS$_2$ flake and $R_0$ is the reference spectrum (see Supplementary information of \cite{roch2019spin} for a detailed description). \\

\noindent{\bf Data availability}\\
The data that support the plots within this paper and other findings of this study are available from the corresponding author upon reasonable request.\\

\noindent \textbf{Acknowledgements} \\
(*) N.L., S.S, I.P.\ and L.S.\ contributed equally to this work. Basel acknowledges funding from the PhD School \textit{Quantum Computing and Quantum Technology}, SNF (Project No.\ 200020\textunderscore 156637), Swiss Nanoscience Institute, and NCCR QSIT. Toulouse acknowledges funding from ANR 2D-vdW-Spin, ANR VallEx, ANR MagicValley, ITN 4PHOTON Marie Sklodowska Curie Grant Agreement No.\ 721394 and the Institut Universitaire de France. K.W. and T.T. acknowledge support from the Elemental Strategy Initiative conducted by the MEXT, Japan, Grant Number JPMXP0112101001,  JSPS KAKENHI Grant Number JP20H00354 and the CREST(JPMJCR15F3), JST. I.C.G.\ thanks the CALMIP initiative for the generous allocation of computational times, through the project p0812, as well as the GENCI-CINES and GENCI-IDRIS for the grant A006096649. We thank Jean-Marie Poumirol for AFM measurements and Jonas~G.~Roch for crucial help at early stages of this work.\\

\noindent \textbf{Author contributions} \\ 
T.G.\ and K.W.\ grew the high quality hBN bulk-crystal. N.L., S.S., I.P.\  and C.R.\ fabricated the encapsulated samples. D.L.\ designed and built the magneto-optics setup (Toulouse). N.L., S.S., I.P., L.S.\ and C.R.\ performed optical spectroscopy measurements. N.L, L.S., I.P., S.S., C.R., A.B.\ and X.M.\ analysed the optical spectra and interpreted the data. I.C.G.\ performed DFT-$GW$-BSE calculations. R.J.W.\ and B.U.\ suggested the experiments and supervised the project. N.L., I.C.G.\ and B.U.\ wrote the manuscript with input from all the authors.\\

\noindent \textbf{Competing interests} \\
The authors declare no competing interests.\\

\noindent{\bf Additional information} \\
Correspondence and requests for materials should be addressed to B.U.
\end{document}